\documentclass[aps,prl,reprint,groupedaddress]{revtex4-1}

\usepackage{amsmath}
\usepackage{graphicx}
\usepackage{enumerate}

\begin{document}
\title{Probing the Origins of Two-State Folding}

\author{Thomas J. Lane}
\affiliation{Department of Chemistry, Stanford University}

\author{Christian R. Schwantes}
\affiliation{Department of Chemistry, Stanford University}

\author{Kyle A. Beauchamp}
\affiliation{Biophysics Program, Stanford University}

\author{Vijay S. Pande}
\affiliation{Department of Chemistry, Biophysics Program, Department of Computer Science, Stanford University}

\date{\today}

\begin{abstract}
Many protein systems fold in a two-state manner. Random models, however, rarely display two-state kinetics and thus such behavior should not be accepted as a default. To date, many theories for the prevalence of two-state kinetics have been presented, but none sufficiently explain the breadth of experimental observations. A model, making a minimum of assumptions, is introduced that suggests two-state behavior is likely for any system with an overwhelmingly populated native state. We show two-state folding is emergent and strengthened by increasing the occupancy population of the native state. Further, the model exhibits a hub-like behavior, with slow interconversions between unfolded states. Despite this, the unfolded state equilibrates quickly relative to the folding time. This apparent paradox is readily understood through this model. Finally, our results compare favorable with experimental measurements of protein folding rates as a function of chain length and $K_{eq}$, and provide new insight into these results.
\end{abstract}

\maketitle

\section{Introduction}

Most small ($< 100$ residues), single domain proteins fold in a two-state manner \cite{Jackson:1991wv, Perl:1998va, Rhoades:2004ii, Barrick:2009bp}. Specifically, protein systems appear to be thermodynamically two-state -- they have only two equilibrium phases (folded and unfolded) -- and also kinetically two-state, exhibiting single exponential kinetics. Protein domains that break this two-state paradigm are usually either large ($> 100$ residues) and folding via one or more intermediates \cite{Bartlett:2009jj}, or small and extremely rapidly folding \cite{Bryngelson:1995hq, Eaton:1999wo, Sabelko:1999wz, GarciaMira:2002ii, Yang:2003di, Yang:2004dz, Yang:2004kb, Gruebele:2005jw, Nguyen:2005ec, Sadqi:2006ia, Liu:2007ij, Liu:2008kj}.

Simple two-state folding kinetics should be considered surprising. Protein chains have a large number of independent conformations available to them, and folding occurs via a stochastic interconversion between these conformations, often described as dynamics evolving on a ``rough'' potential energy function. These dynamics might be thought of as a network, where the nodes are conformations and connections are the rates of interconversion. It is then interesting to ask what the dynamics on a random network looks like \cite{Albert:2002wu, Newman:610163}, and how they compare to protein folding. In such random dynamical systems, two-state behavior is exceptional and rarely seen \footnote{Complexity theory has not traditionally focused on this topic, but the tools are there for its investigation. See, {\it e.g.} \cite{Samukhin:2008hb, Chung:2003el}.}. Later in the paper, we will show why. The fact that two-state behavior is rare in random systems suggests that two-state kinetics cannot be accepted as a default.

Typically two-state kinetics in protein folding is rationalized in terms of a Kramer's rate expression, which postulates the existence of a single dominant free energy barrier between folded and unfolded thermodynamic states. Postulating such a barrier -- in effect, enforcing two-state kinetics -- implies two thermodynamic phases. Here, we investigate the converse; we take two-state thermodynamics as a postulate and show, without ever invoking Kramer's theory or an activated process of any kind, that we can expect such systems to typically be two-state when certain conditions are satisfied \footnote{It should be noted that we do not wish to show this converse holds in general. We suspect that there exist systems that are thermodynamically two-state, but under certain conditions will display multi-exponential kinetics.}. 

One oft-cited explanation is that two-state folding minimizes aggregation, and is therefore evolutionarily advantageous. This argument suggests that folding intermediates are more prone to aggregation than the unfolded or folded state, and therefore biology has attempted to minimize their population during folding \cite{Schindler:1995wr, Jacob:1997tw, Gruebele:2005jw}. While intriguing and certainly possible, there is little direct evidence currently supporting this claim.

\begin{figure}
\includegraphics[width=8cm]{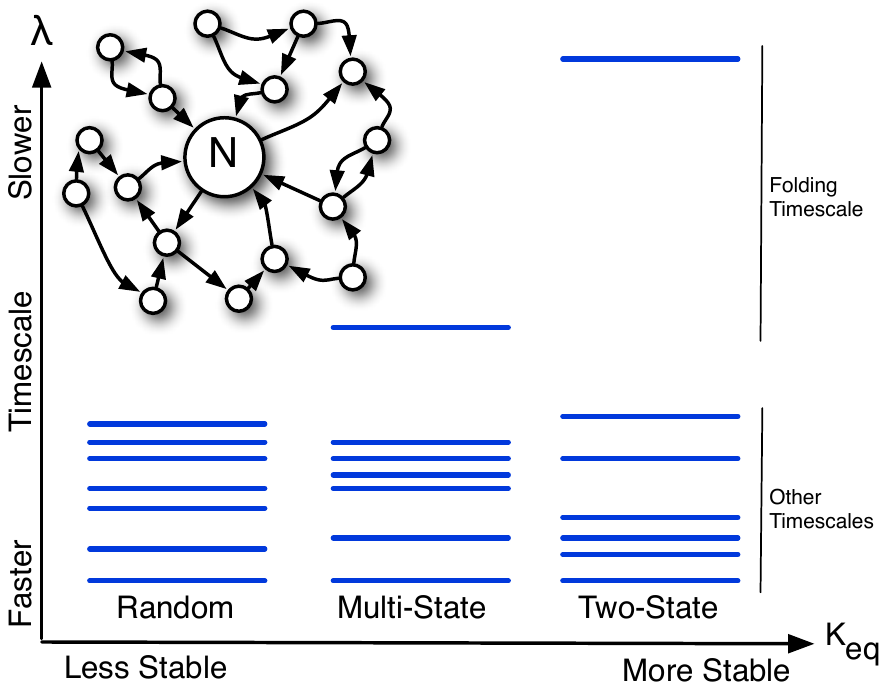}
\caption{Illustration of three different kinetic spectra associated with three protein systems. Each blue horizontal line represents one system timescale (\ref{timetrace}). The theory developed here predicts that as the folded state become more populated, the gap between the slowest and next-slowest timescales will grow. As this separation grows, so will the likelihood that a given experiment will see one dominant relaxation timescale, and classify the system as ``two-state''. These spectra are idealized models that ignore the more complex situations of systems with more than two states. Note that these are \emph{timescale} spectra, representing kinetic processes, and therefore are different from some classic work in folding on thermodynamic energy spectra ({\it e.g.} \cite{Sali:1994vs}). \label{cartoon}}
\end{figure}

An alternative reason for the predominance of two-state folding suggests that most sequences capable of folding are intrinsically two-state. That is, the space of sequences that overwhelmingly populate a native state is enriched in systems that are thermodynamically and kinetically two- or few-state.

We call these competing theories the \emph{aggregation} hypothesis, which dictates that two-state behavior is biologically necessary to avoid aggregation, and the \emph{thermodynamic} hypothesis, dictating two-state folding is intimately linked to folding sequences whether those sequences were formed in the lab or {\it in vivo} \footnote{One additional possibility has been suggested by Robert McGibbon: it could be that two-state systems are considered ``typical'' by the folding community, and non-two state systems are thought of as pathological and therefore not studied. For the purposes of this manuscript we ignore this possibility, but it is important that those working in the field keep it in mind.}. Conclusive proof of either of these, or rejection of both in favor of an alternative, would have a major impact on the folding field. If two-state folding is necessary to avoid aggregation, this certainly has implications for understanding folding \emph{in vivo} and developing therapeutics for Alzheimer's and other aggregation-related diseases. If two-state folding is a physical necessity, then understanding why is a key element in a complete understanding of the protein folding problem.

Here, we present an argument in favor of the thermodynamic hypothesis. We show that for a simple model of protein folding, if there is an overwhelmingly populated native state, two-state thermodynamics and single exponential kinetics are extremely likely. In this model, two-state kinetics are \emph{emergent}, rather than built-in. This allows us to directly analyze the necessary and sufficient conditions for two-state behavior. Interestingly, we find that multi-exponential kinetics can be explained as highly perturbed or relatively unstable two-state systems (Fig.~\ref{cartoon}).

Moreover, the model provides an explanation how two-state systems might produce a ``kinetic hub'' \cite{Bowman:2010hm, Pande:2010hwa, Lane:2012ba}, a dynamical system where most transitions between any two states pass through a third ``hub'' state \cite{Dickson:2012et}. Specifically, we introduce the concept of \emph{slow interconversions with rapid equilibration}, which shows that a situation where the unfolded state equilibrates rapidly does not necessitate a situation where specific structures can reach each other quickly. This seeming paradox is resolved, allowing us to reconcile hub-like kinetics with two-state behavior. The presented model reproduces the bimodal mean first-passage time distribution from all-atom molecular dynamics simulations that originally inspired the hub hypothesis \cite{Bowman:2010fp}, and still retains two-state folding.

\begin{figure}
\includegraphics[width=8cm]{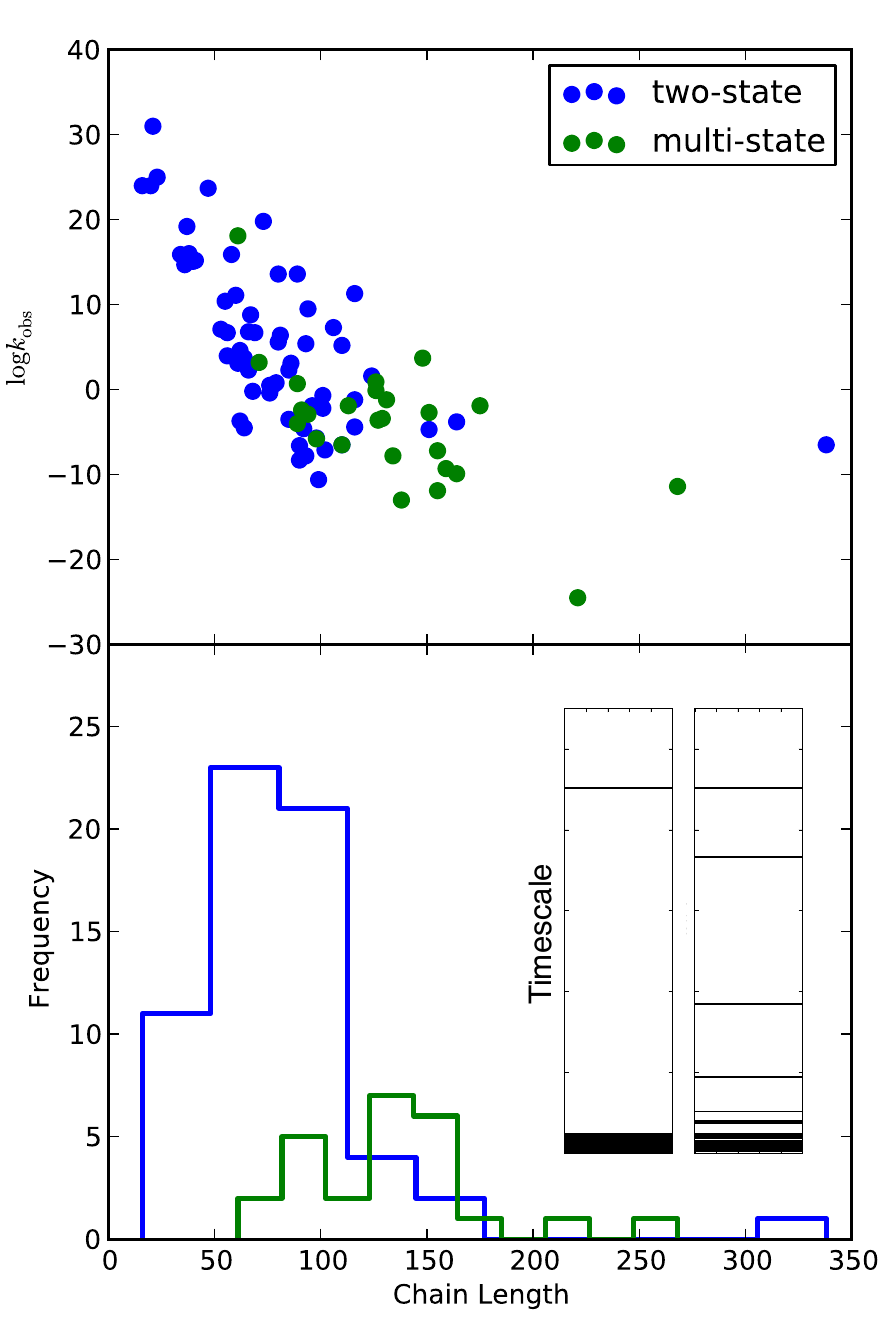}
\caption{Small proteins fold in a two-state manner, while larger proteins exhibit additional timescales. Shown are all proteins in the KineticDB \cite{Bogatyreva:2009jz}, which are classified as two-state (blue) or multi-state (green). Top panel shows folding times as a function of size (units of $k_{\mathrm{obs}}$ are $s^{-1}$), bottom panel shows a histogram of kinetic type as a function of number of residues. At $\sim 100$ residues there is a transition between two-state and multi-state behavior. Insert shows two timescale spectra for the model discussed in the main text ($N=100$ states, $\epsilon = 0.01$, left and $\epsilon = 0.05$, right) showing both two- and higher-state kinetics can be observed in the model. The model predicts more timescales will be observed experimentally as proteins get larger and the perturbation parameter $\epsilon$ increases.  \label{kineticdb}}
\end{figure}

Finally, and crucially, our model provides an explanation for why large proteins (>100 residues) typically exhibit three or more-state folding kinetics (Fig.~\ref{kineticdb}, Fig.~\ref{cv-k3}). The model predicts that as protein systems get larger and larger, additional timescales will be experimentally observable.

In the construction of the model, we postulate only that (A) protein dynamics can be represented by a master equation, (B) the system satisfies detailed balance, and (C) there exists one folded state that is highly populated compared to all other conformations. From these assumptions we build a model that quantitatively contains no additional information, or, equivalently, is the \emph{most random}, using the maximum entropy formalism.

The maximum entropy method is a natural choice for studying folding, because in some sense it reflects the process by which foldable sequences occur. One could conceive of evolution as a random search for a specific target -- the target being a functional structure, and the search occurring through a huge number of possible sequence mutations. The one requirement for this search would be that the functional structure is overwhelmingly populated at equilibrium. Our model shows that \emph{this thermodynamic requirement alone} is sufficient to explain two-state behavior.

\section{Theory of Discrete Time Master Equations}

Our model is framed in the language of Markovian master equations, which give us a framework for implementing the maximum entropy model mentioned. We consider a discrete time propagator $T_{\Delta t}$ that describes the system dynamics. Let $p(t)$ be a function describing the population density of the system at time $t$ -- we are interested in the time evolution of this function. Momentarily we will postulate a partitioning of phase space into $N$ discrete states. In this case, $p(t)$ is a vector, such that $p_i(t)$ is the population of a discrete state $i$ at time $t$. Further, $T_{\Delta t}$ is a stochastic matrix, whose elements $T_{ij}$ describe the probability for the probability density in state $i$ to transfer to state $j$ in the lag time $\Delta t$,
\[
p( n \cdot \Delta t ) = p_0 \> T_{\Delta t}^n 
\]
from initial populations $p_0$. Note that in what follows the specific lag time used will not be too important, so we will drop the $\Delta t$ subscript and just write $T$. 

The system dynamics can then be understood via the eigenmodes of the propagator. The system timescales $\tau_n$ are given by the eigenvalues $\lambda_n$ of $T$,
\begin{equation} \label{implied-timescale}
\tau_n = - \frac{ \Delta t }{ \log( \lambda_n ) }
\end{equation}
while the corresponding eigenvectors describe the exchange of population between states on those timescales. The first eigenvalue is always unity ($\lambda_1 = 1$) corresponding to infinite time. Its corresponding eigenvector is the stationary distribution, denoted $\pi$.

Each row of $T$ is a probability distribution, and therefore must be row-normalized and admits a measure of entropy, loosely speaking the information content of the probability distribution for each row. Further, the entire propagator has an associated entropy (sometimes called the ``caliber'') \cite{JAYNES:1980uh, Stock:2008ii}. Assuming that the distributions described by each row are independent (which is identical to the Markov assumption used to formulate the master equation), we can write the \emph{propagator entropy} as
\begin{equation}\label{entropy}
S_T = - \sum_{i, j} T_{ij} \log T_{ij}
\end{equation}
Maximizing this function, subject to some restraints describing known information, gives the model that makes the fewest assumptions about the system dynamics.

\section{The Maximum Entropy Formalism Allows Thermodynamic Postulates to Result in Kinetic Models}

Using the maximum entropy principle, let us build a model of protein folding. Our goal will be to use well-known facts about proteins as starting assumptions, but limit these as much as possible. Let us postulate:
\begin{enumerate} [A)]

\item Protein dynamics can be described as transitions between $N$ distinct states that partition phase space, where each state is approximately a single conformation (we will call these ``microstates'' states, not to be confused with the thermodynamic folded/unfolded macrostates). We assume dynamics on this space are described by a Markovian propagator $T$. Finally, we expect the number of possible states to be very large (famously estimated by Levinthal to be $\sim 3^{100}$), such that we will not be too hesitant in assuming $N$ is big.

\item The system is ergodic and time-reversible, and therefore the detailed balance condition holds
\begin{equation}\label{detailed-balance}
\pi_i T_{ij} = \pi_j T_{ji}
\end{equation}
where $\pi$ describes the stationary solution to $T$ that is approached asymptotically in time.

\item Proteins have evolved in such a way that there exists a folded state $F$ that has a much larger equilibrium population than all other states, {\it i.e.} $\pi_F \gg \pi_i$ for all $i$ that are not $F$.
\end{enumerate}

Solving for the transition matrix $T$ that maximizes (\ref{entropy}) subject to postulates (A, B, C) is a straightforward exercise in Lagrange multipliers. In what follows, we investigate analytical solutions for the specific case where there is one highly populated native state, and all unfolded states are equally populated at equilibrium. Mathematical detail has been relegated to the supplemental information; we present the results.

Our key result is a timescale spectrum (the eigenspectrum of $T$), which is simply a collection of the system timescales. For instance, in a re-folding experiment a series of exponential decays might be observed in {\it e.g.}~a trace of Trp fluorescence,
\begin{equation}\label{timetrace}
A(t) = \sum_i A_i e^{- t / \tau_i}
\end{equation}
where $A(t)$ is the observable trace, $\tau_i$ are the observed timescales. Finally, the amplitudes $A_i = \langle p_0, \psi^L_i \rangle \langle \mathcal{O},\psi^R_i  \rangle$, with $p_0$ the initial populations of each state when the experiment begins, $\mathcal{O}$ a vector of the average observable value for each state, and $\psi^L_i, \> \psi^R_i$ are the $i^{\mathrm{th}}$ left and right eigenvectors of the propagator, respectively. This amplitude is the mathematical expression of familiar physical concepts. First, the initial populations of each state will affect what experimental response is observed; the factor $\langle p_0, \psi^L_i \rangle$ represents the extent to which a prepared sample (with populations $p_0$) will participate in the mode with timescale $\tau_i$. Second, different experimental probes will report more strongly on certain states than others; the factor $\langle \mathcal{O},\psi^R_i  \rangle$ captures this uneven reporting.

In such an experiment, the ``kinetic spectrum'' is just the collection of $\tau_i$ (Fig.~\ref{cartoon}). There could be one observed exponential, two, or many, depending on the system and experiment. Which timescales are observed will depend on two factors: to be seen, the amplitude $A_i$ must be large enough for a given experiment with limited sensitivity to observe it. Second $\tau_i$ must be in the appropriate range of the experiment's temporal resolution. In this model we compare to experiment by computing the kinetic spectrum, $\{ \tau_i \}$, which is invariant over different experimental probes and initial conditions; we do not compute the amplitudes $\{ A_i \}$, which depend strongly on the experiment under consideration.

\section{A Single Low Energy State Results in A Kinetically and Thermodynamically Two-State System} We take $\pi_F \gg \pi_i$ and $\pi_i = \pi_j$, where $i$ and $j$ are unfolded states. Let's label all such states as $U$, and label the respective populations as $\pi_F$ and $\pi_U$ ($\pi_U$ is the population of just one of $N-1$ unfolded states). With this, it can be shown that $T$ takes the form
\[
T = 
\left( \begin{array}{cccc}
T_{UU}   & T_{UU} & \cdots & T_{UF} \\
T_{UU}  & T_{UU}  &           & T_{UF}     \\
\vdots    &              & \ddots & \vdots  \\
T_{FU}  & T_{FU} & \cdots & T_{FF}
\end{array} \right)
\]
where $T_{UU}$ is the probability of transition from a single unfolded state to any other, $T_{UF}$ is the probability of going from an unfolded to a folded state, and $T_{FU}$ is the probability of the converse. Further, if $\pi_F > \pi_U$, then $T_{FU} < T_{UU} < T_{UF}$; the differences between these get larger as the difference in population between $F$ and $U$ states increases.

This matrix is the foundation of our model. In the next section, we analyze models very near this maximum entropy solution, but where the symmetry in the transition probabilities is broken by a small perturbation; we begin by analyzing the current result, as it represents the simplest case.

%
\begin{figure*}
\includegraphics[width=16cm]{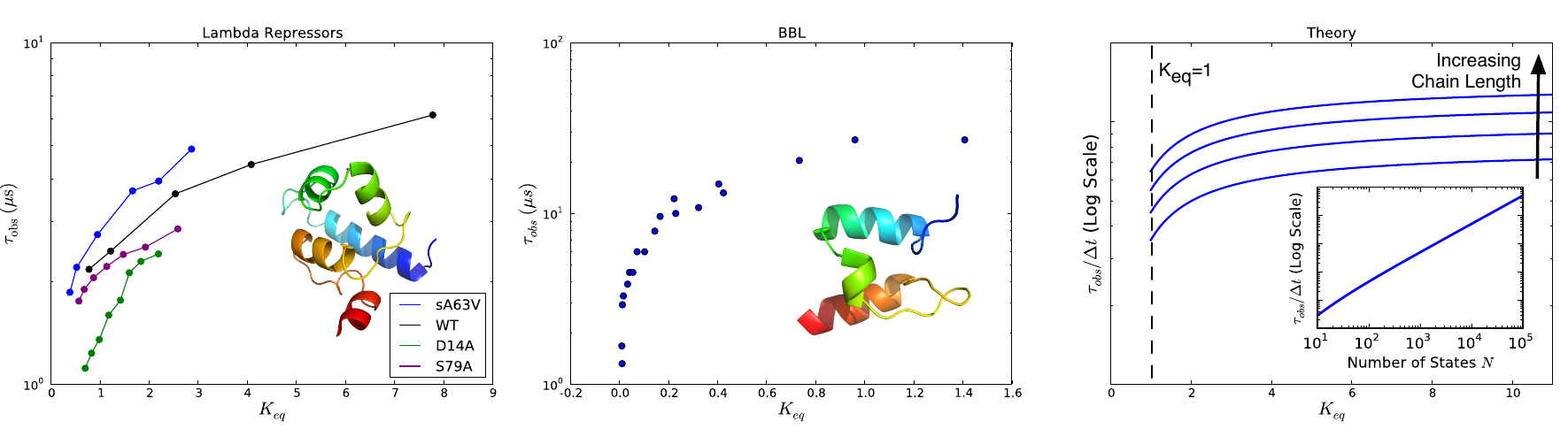}
\caption{A comparison of the theory and experiment: the change in observed folding timescale as a function of $K_{eq}$, modulated experimentally by temperature. Left: a series of lambda repressors from \cite{Yang:2004dz}. Middle: BBL, from \cite{Li:2009bz}. Right: this work (values of $K_{eq} < 1$ not plotted). Since $\tau_3$ is not a function of $K_{eq}$, this plot also demonstrates the gap between $\tau_2$ and $\tau_3$ as a function of $K_{eq}$. Insert: theoretical dependence of folding times on chain length is exponential assuming $R \sim e^{\alpha N}$, which is consistent with experiment (Fig.~\ref{kineticdb}), though is not the only consistent model \cite{Lane:2013ws}. \label{rates-vs-K}}
\end{figure*}

We find three eigenvalues: one stationary ($\lambda_1$), one corresponding to the folding/unfolding reaction ($\lambda_2$), and one corresponding to unfolded state dynamics ($\lambda_3$, with multiplicity $N-2$),
\begin{eqnarray*}
\lambda_1 &=& 1 \\
\lambda_2 &=& 1 - T_{UF} - (N-1) T_{FU} \\
\lambda_3 &=& 1 - T_{UF} - (N-1) T_{UU}
\end{eqnarray*}
showing that since $T_{FU} < T_{UU}$, $\lambda_1 > \lambda_2 > \lambda_3$. Further, for $\pi_F \gg \pi_U$, there will be a significant gap between $\lambda_2$ and $\lambda_3$, leading to a \emph{separation of timescales} consistent with a two-state picture, even though the model consists of $N \gg 2$ states.

The final step of our model involves solving an equation numerically, therefore we cannot write down a closed-form expression for the $\lambda_2$/$\lambda_3$ gap; it is possible, however,
to plot this timescale gap as a function of $K_{eq}$ or the number of configuration states $N$. Figure \ref{rates-vs-K} shows the scaling of the folding timescale $\tau_2 = - \Delta t / \log(\lambda_2)$ with $K_{eq}$, and compares this scaling with two experimental systems, lambda repressor \cite{Yang:2004dz} and BBL \cite{Li:2009bz}. Further, $\lambda_3$ is not a function of $K_{eq}$ in this model. Thus, the rightmost panel of Fig.~\ref{rates-vs-K} demonstrates the theoretical scaling of the gap between the slowest (folding) timescale and the next-slowest system timescale. A precise treatment of this second-slowest timescale is performed in the next section.

It is critical to note that the discussion of modulating $K_{eq}$ is restricted to changing the conditions for a \emph{single protein sequence}. Our model contains no features that allow it to distinguish between sequences. Thus, we use the term ``stability'' to mean the relative population of the native state between under two different conditions, and not relative stability between mutants or different proteins (that might be compared via a $\Delta G$ of folding or melting temperature $T_m$).

Why is a gap between $\lambda_2$ and $\lambda_3$ a signature of two-state kinetics? In a system where the folding timescale is much slower than the rest, an experiment designed to study folding may be poorly suited (too low-resolution) to measure faster kinetics. Further, probes designed specifically to study folding may have a large amplitude response to folding kinetics, but not to other kinetic modes in the system. Mathematically, if $\langle \mathcal{O},\psi^R_2  \rangle$ is maximized ($\psi^R_2$ is the eigenvector describing folding), then $\langle \mathcal{O},\psi^R_n \rangle \ n \geq 3$ will be small, since the eigenvectors $\psi^R_i$ are orthogonal. Therefore, experiments well-designed to study folding will measure other system modes at a much lower amplitude.

Recently, as higher resolution instrumentation has been developed \cite{Prigozhin:2013ex}, faster timescales such as these have been found in folding systems that were previously considered two-state. Perhaps the best single example of this is the villin headpiece. In \cite{Wang:2003ha}, the kinetics of villin were measured by NMR lineshape analysis and fit well to a two-state model with a folding time of order $\mu$s. Later, laser-induced temperature jump and ultra-fast triplet-triplet energy transfer experiments reveled additional dynamical processes, including intermediate formation at 70 ns \cite{Kubelka:2006bv} and native state locking/unlocking at 170 ns \cite{Reiner:2010bp}. Thus, the villin timescale spectrum has one relatively slow mode, its folding time, and at least two faster modes. This is consistent with our model, which predicts proteins have one slow mode ($\lambda_2$) and a number of faster modes ($\lambda_n, 3 \leq \ n \leq N$).

In systems where the separation between slow and fast modes is very large -- larger than the separation for villin, for instance -- it is likely that an experiment designed to measure the slow folding timescale will not measure faster timescales. Thus, only one kinetic timescale is ever seen experimentally, and we call such systems kinetically ``two-state''.

\begin{figure}
\includegraphics[width=8cm]{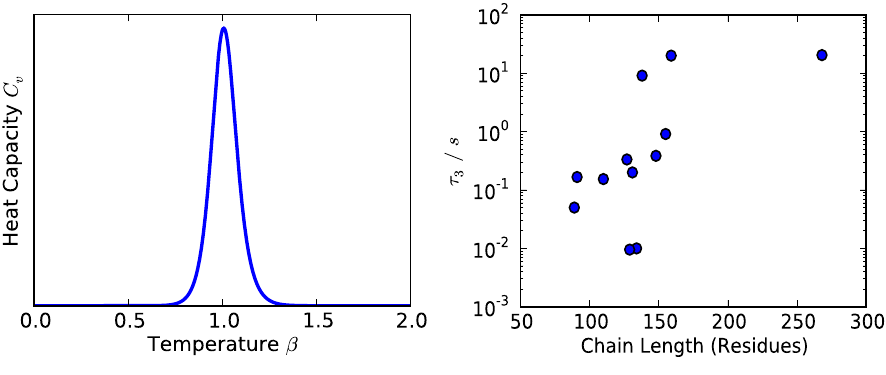}
\caption{Left: the calculated $C_V$ vs $\beta$ curve (for $N=100$), showing a phase transition at the melting temperature  $\beta = 1$. Right: experimental dependence of $\tau_3$ with protein chain length $R$. Data from 12 multi-state proteins reported in the KineticDB \cite{Bogatyreva:2009jz}. \label{cv-k3}}
\end{figure}

We have observed the model is kinetically two-state, but it is also (by construction, due to the choice of $\pi$) thermodynamically two-state, as demonstrated by a sharply peaked $C_V$ curve. Denote the free energy of each state by $A_U$ for unfolded and $A_F$ for the folded states. Then set our scale of energy such that $A_U = 0$. With this, we can write the ratio of state populations
\[
\frac{\pi_F}{\pi_U} = \frac{ e^{-\beta A_F} }{ e^0 } = e^{-\beta A_F}
\]
now, scale our units of temperature such that the folding temperature, where $\pi_F = 0.5$, occurs at $\beta = 1$. Then it is clear from the above that $A_F = - \log( N-1 )$. The partition function is
\[
Z = \sum_i e^{-\beta A_i} = (N-1) + (N-1)^\beta
\]
resulting in the heat capacity
\[
C_V = \beta^2 (N-1)^{\beta+1} \left[ \frac{  \log (N-1) }{ (N-1) + (N-1)^{\beta} } \right]^2
\]
which exhibits a first order phase transition at $\beta=1$, consistent with what is observed in experiment (Fig. \ref{cv-k3}).
%

\section{Perturbation of the Model Shows Two-State Folding is Robust}

While this simple model demonstrates a minimal set of sufficient requirements for two-state folding, it retains an artificial symmetry -- all the rates in each set $\{T_{UU} \}, \{ T_{UF} \}, \{ T_{FU} \}$ are identical. Such symmetry can be broken by adding random ``noise'' to the transition matrix elements. Robust two-state folding should not be affected by such a perturbation -- experimentally, a single mutation or slight change in experimental conditions is insufficient to disrupt two-state behavior in the majority of systems.

A reasonable perturbation is the addition of a random Gaussian to each element of $T$, in the form of a matrix $T'$ whose elements are derived from Gaussians
\[
\tilde{T} = T + \epsilon T'
\]
where primes to denote perturbing terms, tildes the resulting perturbed solution, and $\epsilon$ is a control parameter denoting the size of the perturbation \footnote{We have only carried out the perturbation to first order. The perturbations we want to consider are small, since large perturbations stray too far from the entropy maximum solution and therefore are expected to imply work done by biology.}. In the supplemental information, we show that one can construct such a perturbation while ensuring $\tilde{T}$ maintains detailed balance and is a stochastic matrix, in the thermodynamic limit ($N \to \infty$). The perturbation derives from a symmetric matrix whose elements are drawn from a Gaussian distribution -- such a matrix is known as a member of the Gaussian orthogonal ensemble (GOE) \cite{Mehta:2004wq}. 
%
\begin{figure}
\includegraphics[width=8cm]{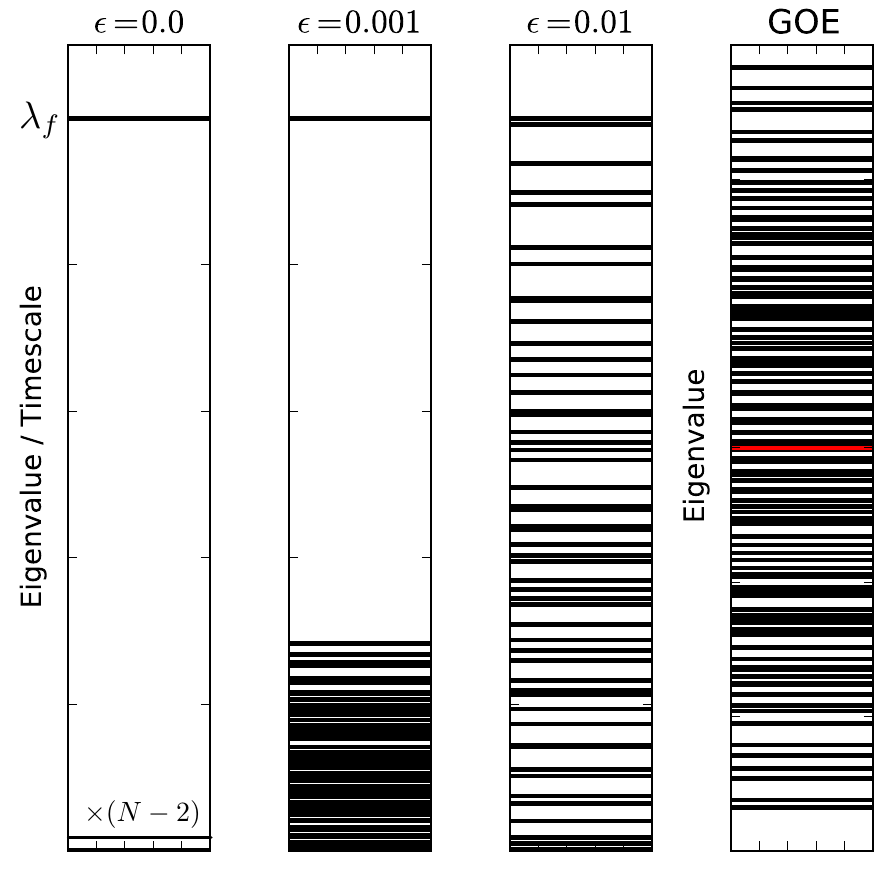}
\caption{The model's eigenspectra ($\{ \lambda_i \}$) under no perturbation (left), a small perturbation (middle left), and a large perturbation (middle right). These plots are for a single instance of the distribution indicated in the main paper, with $\epsilon$ set as indicated. The far right is the spectrum of one sample from the GOE (red line is origin). In the absence of any native bias ($\pi_F \approx \pi_U$), the kinetic spectrum would be qualitatively similar to just the GOE (with all of the eigenvalues positive). Parameters were $N=100$ and $K_{eq} = 1.0$. \label{spectra}}
\end{figure}

From this, we find the probability density of obtaining a spectrum $P(\tilde{\lambda_n})$ under a random perturbation is
\[
P(\tilde{\lambda_n}) = C \prod_{n=3}^N e^{- \tilde{ \lambda_n}^2 / 2 \epsilon^2 } \sum_{3 \geq i}^N \sum_{ j > i}^N | \tilde{\lambda_i} - \tilde{\lambda_j} |
     \]
where we have considered only the eigenvalues representing dynamics in the unfolded state ($\lambda_3$). Here $C$ is simply a constant that normalizes the distribution.

This perturbation has the effect of splitting the degeneracy and spreading out the previously overlapping eigenvalues (Fig.~\ref{spectra}). One can see this by noticing that the terms of form $| \tilde{\lambda_i} - \tilde{\lambda_j} |$ require the probability density go to zero as two eigenvalues get close together. This effect is known as \emph{level repulsion} in random matrix theory.

What consequences does this perturbation have for protein folding? One can see that for small perturbations, the symmetry of the original degenerate model is broken, but a spectral gap between the unfolded timescale and the folding timescale ($\lambda_2$) still exists (Fig. \ref{spectra}) \footnote{Because of the spreading of the eigenvalues, this gap will be smaller than before. One could calculate the probabilistic difference from the Tracy-Widom distribution \cite{Tracy:1996uc}, but we have not undertaken this task (yet).}. Once perturbations get very large, however, one expects that the timescales of unfolded state dynamics will spread sufficiently to be comparable to the folding timescale. This will destroy the two-state features of the model, and shows that, as stated in the introduction, highly random models will not exhibit two-state behavior. We conclude that while two-state folding for this model is relatively robust, under server perturbations multi-exponential, non-two state kinetics may arise.

\section{Three or Greater State Behavior May Be Observed, Especially in Large Systems.}

Random matrix theory also provides an estimate for the relative timescale of the slowest non-folding process,
\[
\frac{\tau_3}{\Delta t} = - \left[ {\frac{\alpha}{2} R + \log \epsilon \sqrt{2} } \right]^{-1}
\]
as a function of the size of the random perturbation $\epsilon$ and chain length $R$, assuming the exponential scaling $N \sim e^{\alpha R}$ \footnote{Note that this expression requires $\epsilon \sim \mathcal{O}(1/N)$, which for large state number $N$ makes $\epsilon$ very small. The choice of $\epsilon$, however, is arbitrary; it is simply a control parameter.}. While the expression for $\tau_3$ contains too many unknown parameters ($\epsilon$, $\alpha$, $\Delta t$) to make meaningful quantitative predictions of experiment, it does suggest that the slowest non-folding timescales in proteins should increase with chain length. This is consistent with what is seen experimentally for the 12 multi-state proteins for which the slowest non-folding timescale is reported in the literature (Fig.~\ref{cv-k3}) \cite{Bogatyreva:2009jz} -- additional data will be necessary to definitively confirm this prediction.

\section{The Model Exhibits Native Hub-Like Behavior}

\begin{figure}
\includegraphics[width=8cm]{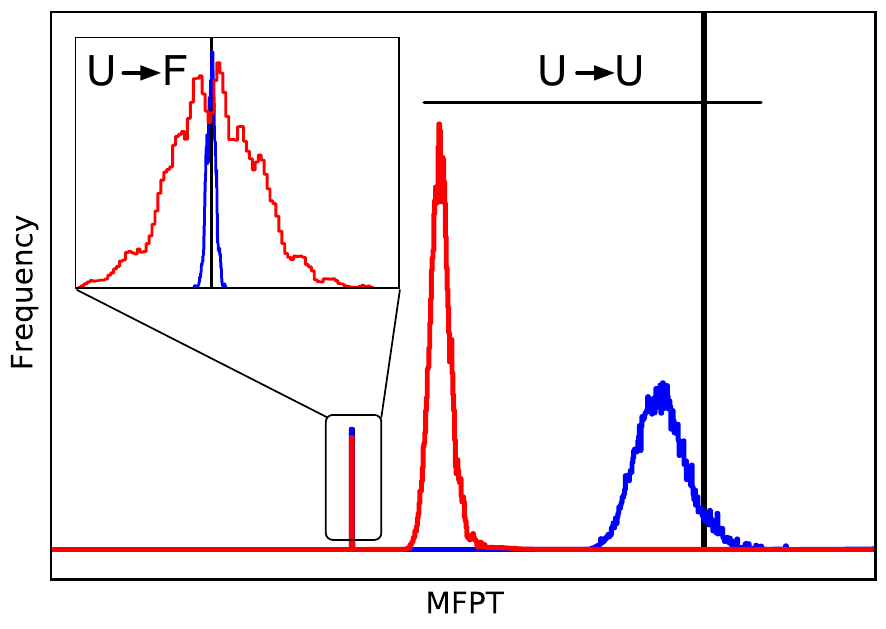}
\caption{MFPT distribution (normalized) for an unperturbed (black, $\delta$-functions), slightly perturbed (blue, $\epsilon = 10^{-5}$) and significantly perturbed (red, $\epsilon = 10^{-4}$) models. The distribution is bimodal, or ``hub-like''. Shown are 1000 numerical samples for $N=100$ and $K_{eq} = 1.0$. There are some large MFPTs representing $F \to U$ transitions that are not shown here. \label{mfpt}}
\end{figure}

The mean first passage time (MFPT) is the expected time it takes for a walker starting at state $i$ of a Markov chain to reach state $j$ for the first time. It is apparent from inspection that, due to the fact that $T_{UU} < T_{UF}$, we expect the MFPT from an unfolded state to any other unfolded state to be \emph{slower} than the passage time from that state to the native state. A plot of the distributions of MFPTs from every state to every other is therefore bimodal (Fig.~\ref{mfpt}), a property that has been described as ``hub-like'' \cite{Bowman:2010fp} after it was witnessed in all-atom molecular dynamics simulations \cite{Rao:2004ce, Bowman:2010hm, Lane:2011fp} (recently more sensitive measures of hub-like phenomenology have been proposed and employed a model very similar to this one \cite{Dickson:2012et, Dickson:2013cf}). Numerical simulations show a small perturbation spreads out such a distribution, but does not destroy the bimodality (Fig.~\ref{mfpt}). The system only exhibits these hub-like behaviors when there is a single native state, $\pi_F \gg \{ \pi_U \}$.

Slow MFPTs between unfolded states might seem inconsistent with the fact that $\lambda_3$, the eigenvalues corresponding to the unfolded state dynamics, represent timescales that are much faster than the folding time. This ``paradox'' arises because the eigenvalue ($\lambda_3$) is a measure of the \emph{ensemble} dynamics of the system, while MFPTs measure dynamics at the level of a single protein visiting specific states. Imagine the following two cases. If we monitor a single protein molecule as it folds, it will visit many unfolded states before folding. However, the chances that it reaches \emph{one particular} unfolded state before folding is very small -- it is much more likely to visit the native state before this single unfolded state. Quantitatively, this results in a smaller MFPT for $U \to F$ transitions than $U \to U$ transitions (Fig.~\ref{mfpt}). 

Next, consider an ensemble of proteins. Because they each visit a large number of unfolded states before reaching the single native state, they are able to spread out quickly, equilibrating all of the unfolded states, before these unfolded states have a chance to equilibrate with the native state. This results in a system that exhibits \emph{slow interconversions with rapid equilibration}.

This phenomenon is purely a result of dividing the unfolded state into many parts -- in effect, increasing the resolution of non-native dynamics. Our model, being phenomenological, cannot address whether or not significant energetic or enthalpic barriers exist in the non-native ensemble. This is an important outstanding question that will require further work to address. It is important to note that, at high spatial resolution, hub-like kinetics might be present purely due to the size of the unfolded state space, regardless of if such barriers exist or not. A corollary of this is that at low resolution, this hub-like behavior will disappear -- in the limit of two states, it is by definition impossible to have any kind of network hub.

This interpretation is consistent not only with traditional views of rapid unfolded-state equilibration, but also recent reports of relatively slow interconversions between non-native conformations \cite{Waldauer:2010vg, Bowman:2010fp, Pande:2010hw, Lane:2011fp}. This model provides a lens for reconciling these views.

\section{Conclusions}

What are the minimal sufficient features for a folding sequence to exhibit two-state kinetics and thermodynamics? Is two-state folding biologically advantageous, or a physical requirement?  The model presented here suggests that simply a large enough energy gap between folded and unfolded states is enough to result in two-state behavior. 

Further, this model
\begin{itemize}

\item Explains why two-state systems are common in small proteins, but additional timescales appear in larger proteins. The model attributes this to the nature of protein thermodynamics, and does not invoke an aggregation-based evolutionary hypothesis.

\item Shows why additional fast timescales, usually not directly involved in folding, can be observed in traditionally two-state systems such as villin. 

\item Displays two-state kinetics without making reference to or assuming an activated process. Agnosticism in this regard allows us to analyze non-native dynamics in ways that models that begin with an activated process cannot.

\item Reconciles the hub-like kinetics observed in simulation and with two-state kinetics, through the concept of rapid equilibration with slow interconversions. 

\item Proposes a new interpretation of multi-exponential kinetics in fast folding proteins, and this mechanism is seen to be in agreement with reported experiments. The model makes clear predictions about how folding times change with respect to $K_{eq}$.

\item Introduced an approach based on maximizing the entropy of a dynamical propagator given known information as a way of probing protein folding theoretically. More sophisticated models, that include more detailed structural information and precise state energy structures, may yield additional experimental predictions.

\end{itemize}
Each of these items provides either new insight into empirical observations made in experiments or simulations that were previously poorly understood, or pushes the methods  employed currently in the construction of analytical theories of protein folding.

This model brings into focus two significant research questions that remain unresolved. First, this model says little about what the topologies of realistic propagators of protein dynamics look like, and to what extent those topologies dictate protein dynamics. Second, we have so far been unable to address the nature of dynamics in the unfolded state. Whether or not these non-native dynamics are restricted by significant barriers remains an open question, one that requires a microscopic theory (as opposed to the phenomenological theory presented here) validated careful experimentation and simulation to fully understand.

\section{Supplemental Information}
Supplementary information provides mathematical detail. It includes the Lagrange multiplier-based solution of the  maximum entropy propagator $T$, an analysis of the eigenvalues and eigenvectors of that propagator, a perturbation-theoretic approach to the stability of that eigenstructure, and a calculation of the timescale gap between $\lambda_2$ and $\lambda_3$. The results of these calculations were presented in the main paper.

\section{acknowledgments}
TJL would like to acknowledge a conversation with Attila Szabo on the reconciliation of hub-like and two-state kinetics explained here. This manuscript has benefited greatly from discussions with Max Prigozhin, Hannah Gelman, Alex Grosberg, and Geoff Rollins. TJL was supported by an NSF GRF. Further support was provided by the NSF (NSF-MCB-0954714) and NIH (R01-GM062868).

\bibliography{papers2.bib}

\begin{thebibliography}{51}%
\makeatletter
\providecommand \@ifxundefined [1]{%
 \@ifx{#1\undefined}
}%
\providecommand \@ifnum [1]{%
 \ifnum #1\expandafter \@firstoftwo
 \else \expandafter \@secondoftwo
 \fi
}%
\providecommand \@ifx [1]{%
 \ifx #1\expandafter \@firstoftwo
 \else \expandafter \@secondoftwo
 \fi
}%
\providecommand \natexlab [1]{#1}%
\providecommand \enquote  [1]{``#1''}%
\providecommand \bibnamefont  [1]{#1}%
\providecommand \bibfnamefont [1]{#1}%
\providecommand \citenamefont [1]{#1}%
\providecommand \href@noop [0]{\@secondoftwo}%
\providecommand \href [0]{\begingroup \@sanitize@url \@href}%
\providecommand \@href[1]{\@@startlink{#1}\@@href}%
\providecommand \@@href[1]{\endgroup#1\@@endlink}%
\providecommand \@sanitize@url [0]{\catcode `\\12\catcode `\$12\catcode
  `\&12\catcode `\#12\catcode `\^12\catcode `\_12\catcode `\%12\relax}%
\providecommand \@@startlink[1]{}%
\providecommand \@@endlink[0]{}%
\providecommand \url  [0]{\begingroup\@sanitize@url \@url }%
\providecommand \@url [1]{\endgroup\@href {#1}{\urlprefix }}%
\providecommand \urlprefix  [0]{URL }%
\providecommand \Eprint [0]{\href }%
\providecommand \doibase [0]{http://dx.doi.org/}%
\providecommand \selectlanguage [0]{\@gobble}%
\providecommand \bibinfo  [0]{\@secondoftwo}%
\providecommand \bibfield  [0]{\@secondoftwo}%
\providecommand \translation [1]{[#1]}%
\providecommand \BibitemOpen [0]{}%
\providecommand \bibitemStop [0]{}%
\providecommand \bibitemNoStop [0]{.\EOS\space}%
\providecommand \EOS [0]{\spacefactor3000\relax}%
\providecommand \BibitemShut  [1]{\csname bibitem#1\endcsname}%
\let\auto@bib@innerbib\@empty
\bibitem [{\citenamefont {Jackson}\ and\ \citenamefont
  {Fersht}(1991)}]{Jackson:1991wv}%
  \BibitemOpen
  \bibfield  {author} {\bibinfo {author} {\bibfnamefont {S.}~\bibnamefont
  {Jackson}}\ and\ \bibinfo {author} {\bibfnamefont {A.~R.}\ \bibnamefont
  {Fersht}},\ }\href@noop {} {\bibfield  {journal} {\bibinfo  {journal}
  {Biochemistry}\ }\textbf {\bibinfo {volume} {30}},\ \bibinfo {pages} {10428}
  (\bibinfo {year} {1991})}\BibitemShut {NoStop}%
\bibitem [{\citenamefont {Perl}\ \emph {et~al.}(1998)\citenamefont {Perl},
  \citenamefont {Welker}, \citenamefont {Schindler}, \citenamefont
  {Schr{\"o}der}, \citenamefont {Marahiel}, \citenamefont {Jaenicke},\ and\
  \citenamefont {Schmid}}]{Perl:1998va}%
  \BibitemOpen
  \bibfield  {author} {\bibinfo {author} {\bibfnamefont {D.}~\bibnamefont
  {Perl}}, \bibinfo {author} {\bibfnamefont {C.}~\bibnamefont {Welker}},
  \bibinfo {author} {\bibfnamefont {T.}~\bibnamefont {Schindler}}, \bibinfo
  {author} {\bibfnamefont {K.}~\bibnamefont {Schr{\"o}der}}, \bibinfo {author}
  {\bibfnamefont {M.~A.}\ \bibnamefont {Marahiel}}, \bibinfo {author}
  {\bibfnamefont {R.}~\bibnamefont {Jaenicke}}, \ and\ \bibinfo {author}
  {\bibfnamefont {F.~X.}\ \bibnamefont {Schmid}},\ }\href@noop {} {\bibfield
  {journal} {\bibinfo  {journal} {Nat. Struct. Biol.}\ }\textbf {\bibinfo
  {volume} {5}},\ \bibinfo {pages} {229} (\bibinfo {year} {1998})}\BibitemShut
  {NoStop}%
\bibitem [{\citenamefont {Rhoades}\ \emph {et~al.}(2004)\citenamefont
  {Rhoades}, \citenamefont {Cohen}, \citenamefont {Schuler},\ and\
  \citenamefont {Haran}}]{Rhoades:2004ii}%
  \BibitemOpen
  \bibfield  {author} {\bibinfo {author} {\bibfnamefont {E.}~\bibnamefont
  {Rhoades}}, \bibinfo {author} {\bibfnamefont {M.}~\bibnamefont {Cohen}},
  \bibinfo {author} {\bibfnamefont {B.}~\bibnamefont {Schuler}}, \ and\
  \bibinfo {author} {\bibfnamefont {G.}~\bibnamefont {Haran}},\ }\href@noop {}
  {\bibfield  {journal} {\bibinfo  {journal} {J. Am. Chem. Soc.}\ }\textbf
  {\bibinfo {volume} {126}},\ \bibinfo {pages} {14686} (\bibinfo {year}
  {2004})}\BibitemShut {NoStop}%
\bibitem [{\citenamefont {Barrick}(2009)}]{Barrick:2009bp}%
  \BibitemOpen
  \bibfield  {author} {\bibinfo {author} {\bibfnamefont {D.}~\bibnamefont
  {Barrick}},\ }\href@noop {} {\bibfield  {journal} {\bibinfo  {journal} {Phys.
  Biol.}\ }\textbf {\bibinfo {volume} {6}},\ \bibinfo {pages} {015001}
  (\bibinfo {year} {2009})}\BibitemShut {NoStop}%
\bibitem [{\citenamefont {Bartlett}\ and\ \citenamefont
  {Radford}(2009)}]{Bartlett:2009jj}%
  \BibitemOpen
  \bibfield  {author} {\bibinfo {author} {\bibfnamefont {A.~I.}\ \bibnamefont
  {Bartlett}}\ and\ \bibinfo {author} {\bibfnamefont {S.~E.}\ \bibnamefont
  {Radford}},\ }\href@noop {} {\bibfield  {journal} {\bibinfo  {journal} {Nat
  Struct Mol Biol}\ }\textbf {\bibinfo {volume} {16}},\ \bibinfo {pages} {582}
  (\bibinfo {year} {2009})}\BibitemShut {NoStop}%
\bibitem [{\citenamefont {Bryngelson}\ \emph {et~al.}(1995)\citenamefont
  {Bryngelson}, \citenamefont {Onuchic}, \citenamefont {Socci},\ and\
  \citenamefont {Wolynes}}]{Bryngelson:1995hq}%
  \BibitemOpen
  \bibfield  {author} {\bibinfo {author} {\bibfnamefont {J.~D.}\ \bibnamefont
  {Bryngelson}}, \bibinfo {author} {\bibfnamefont {J.}~\bibnamefont {Onuchic}},
  \bibinfo {author} {\bibfnamefont {N.~D.}\ \bibnamefont {Socci}}, \ and\
  \bibinfo {author} {\bibfnamefont {P.}~\bibnamefont {Wolynes}},\ }\href@noop
  {} {\bibfield  {journal} {\bibinfo  {journal} {Proteins}\ }\textbf {\bibinfo
  {volume} {21}},\ \bibinfo {pages} {167} (\bibinfo {year} {1995})}\BibitemShut
  {NoStop}%
\bibitem [{\citenamefont {Eaton}(1999)}]{Eaton:1999wo}%
  \BibitemOpen
  \bibfield  {author} {\bibinfo {author} {\bibfnamefont {W.}~\bibnamefont
  {Eaton}},\ }\href@noop {} {\bibfield  {journal} {\bibinfo  {journal} {Proc
  Natl Acad Sci USA}\ }\textbf {\bibinfo {volume} {96}},\ \bibinfo {pages}
  {5897} (\bibinfo {year} {1999})}\BibitemShut {NoStop}%
\bibitem [{\citenamefont {Sabelko}\ \emph {et~al.}(1999)\citenamefont
  {Sabelko}, \citenamefont {Ervin},\ and\ \citenamefont
  {Gruebele}}]{Sabelko:1999wz}%
  \BibitemOpen
  \bibfield  {author} {\bibinfo {author} {\bibfnamefont {J.}~\bibnamefont
  {Sabelko}}, \bibinfo {author} {\bibfnamefont {J.}~\bibnamefont {Ervin}}, \
  and\ \bibinfo {author} {\bibfnamefont {M.}~\bibnamefont {Gruebele}},\
  }\href@noop {} {\bibfield  {journal} {\bibinfo  {journal} {Proceedings of the
  National Academy of Sciences}\ }\textbf {\bibinfo {volume} {96}},\ \bibinfo
  {pages} {6031} (\bibinfo {year} {1999})}\BibitemShut {NoStop}%
\bibitem [{\citenamefont {Garcia-Mira}\ \emph {et~al.}(2002)\citenamefont
  {Garcia-Mira}, \citenamefont {Sadqi}, \citenamefont {Fischer}, \citenamefont
  {Sanchez-Ruiz},\ and\ \citenamefont {Mu{\~n}oz}}]{GarciaMira:2002ii}%
  \BibitemOpen
  \bibfield  {author} {\bibinfo {author} {\bibfnamefont {M.~M.}\ \bibnamefont
  {Garcia-Mira}}, \bibinfo {author} {\bibfnamefont {M.}~\bibnamefont {Sadqi}},
  \bibinfo {author} {\bibfnamefont {N.}~\bibnamefont {Fischer}}, \bibinfo
  {author} {\bibfnamefont {J.~M.}\ \bibnamefont {Sanchez-Ruiz}}, \ and\
  \bibinfo {author} {\bibfnamefont {V.}~\bibnamefont {Mu{\~n}oz}},\ }\href@noop
  {} {\bibfield  {journal} {\bibinfo  {journal} {Science}\ }\textbf {\bibinfo
  {volume} {298}},\ \bibinfo {pages} {2191} (\bibinfo {year}
  {2002})}\BibitemShut {NoStop}%
\bibitem [{\citenamefont {Yang}\ and\ \citenamefont
  {Gruebele}(2003)}]{Yang:2003di}%
  \BibitemOpen
  \bibfield  {author} {\bibinfo {author} {\bibfnamefont {W.~Y.}\ \bibnamefont
  {Yang}}\ and\ \bibinfo {author} {\bibfnamefont {M.}~\bibnamefont
  {Gruebele}},\ }\href@noop {} {\bibfield  {journal} {\bibinfo  {journal}
  {Nature}\ }\textbf {\bibinfo {volume} {423}},\ \bibinfo {pages} {193}
  (\bibinfo {year} {2003})}\BibitemShut {NoStop}%
\bibitem [{\citenamefont {Yang}\ and\ \citenamefont
  {Gruebele}(2004{\natexlab{a}})}]{Yang:2004dz}%
  \BibitemOpen
  \bibfield  {author} {\bibinfo {author} {\bibfnamefont {W.~Y.}\ \bibnamefont
  {Yang}}\ and\ \bibinfo {author} {\bibfnamefont {M.}~\bibnamefont
  {Gruebele}},\ }\href@noop {} {\bibfield  {journal} {\bibinfo  {journal}
  {Biochemistry}\ }\textbf {\bibinfo {volume} {43}},\ \bibinfo {pages} {13018}
  (\bibinfo {year} {2004}{\natexlab{a}})}\BibitemShut {NoStop}%
\bibitem [{\citenamefont {Yang}\ and\ \citenamefont
  {Gruebele}(2004{\natexlab{b}})}]{Yang:2004kb}%
  \BibitemOpen
  \bibfield  {author} {\bibinfo {author} {\bibfnamefont {W.~Y.}\ \bibnamefont
  {Yang}}\ and\ \bibinfo {author} {\bibfnamefont {M.}~\bibnamefont
  {Gruebele}},\ }\href@noop {} {\bibfield  {journal} {\bibinfo  {journal}
  {Biophysj}\ }\textbf {\bibinfo {volume} {87}},\ \bibinfo {pages} {596}
  (\bibinfo {year} {2004}{\natexlab{b}})}\BibitemShut {NoStop}%
\bibitem [{\citenamefont {Gruebele}(2005)}]{Gruebele:2005jw}%
  \BibitemOpen
  \bibfield  {author} {\bibinfo {author} {\bibfnamefont {M.}~\bibnamefont
  {Gruebele}},\ }\href@noop {} {\bibfield  {journal} {\bibinfo  {journal}
  {Comptes Rendus Biologies}\ }\textbf {\bibinfo {volume} {328}},\ \bibinfo
  {pages} {701} (\bibinfo {year} {2005})}\BibitemShut {NoStop}%
\bibitem [{\citenamefont {Nguyen}\ \emph {et~al.}(2005)\citenamefont {Nguyen},
  \citenamefont {Jager}, \citenamefont {Kelly},\ and\ \citenamefont
  {Gruebele}}]{Nguyen:2005ec}%
  \BibitemOpen
  \bibfield  {author} {\bibinfo {author} {\bibfnamefont {H.}~\bibnamefont
  {Nguyen}}, \bibinfo {author} {\bibfnamefont {M.}~\bibnamefont {Jager}},
  \bibinfo {author} {\bibfnamefont {J.~W.}\ \bibnamefont {Kelly}}, \ and\
  \bibinfo {author} {\bibfnamefont {M.}~\bibnamefont {Gruebele}},\ }\href@noop
  {} {\bibfield  {journal} {\bibinfo  {journal} {J. Phys. Chem. B}\ }\textbf
  {\bibinfo {volume} {109}},\ \bibinfo {pages} {15182} (\bibinfo {year}
  {2005})}\BibitemShut {NoStop}%
\bibitem [{\citenamefont {Sadqi}\ \emph {et~al.}(2006)\citenamefont {Sadqi},
  \citenamefont {Fushman},\ and\ \citenamefont {Mu{\~n}oz}}]{Sadqi:2006ia}%
  \BibitemOpen
  \bibfield  {author} {\bibinfo {author} {\bibfnamefont {M.}~\bibnamefont
  {Sadqi}}, \bibinfo {author} {\bibfnamefont {D.}~\bibnamefont {Fushman}}, \
  and\ \bibinfo {author} {\bibfnamefont {V.}~\bibnamefont {Mu{\~n}oz}},\
  }\href@noop {} {\bibfield  {journal} {\bibinfo  {journal} {Nature}\ }\textbf
  {\bibinfo {volume} {442}},\ \bibinfo {pages} {317} (\bibinfo {year}
  {2006})}\BibitemShut {NoStop}%
\bibitem [{\citenamefont {Liu}\ and\ \citenamefont
  {Gruebele}(2007)}]{Liu:2007ij}%
  \BibitemOpen
  \bibfield  {author} {\bibinfo {author} {\bibfnamefont {F.}~\bibnamefont
  {Liu}}\ and\ \bibinfo {author} {\bibfnamefont {M.}~\bibnamefont {Gruebele}},\
  }\href@noop {} {\bibfield  {journal} {\bibinfo  {journal} {Journal of
  Molecular Biology}\ }\textbf {\bibinfo {volume} {370}},\ \bibinfo {pages}
  {574} (\bibinfo {year} {2007})}\BibitemShut {NoStop}%
\bibitem [{\citenamefont {Liu}\ \emph {et~al.}(2008)\citenamefont {Liu},
  \citenamefont {Du}, \citenamefont {Fuller}, \citenamefont {Davoren},
  \citenamefont {Wipf}, \citenamefont {Kelly},\ and\ \citenamefont
  {Gruebele}}]{Liu:2008kj}%
  \BibitemOpen
  \bibfield  {author} {\bibinfo {author} {\bibfnamefont {F.}~\bibnamefont
  {Liu}}, \bibinfo {author} {\bibfnamefont {D.}~\bibnamefont {Du}}, \bibinfo
  {author} {\bibfnamefont {A.~A.}\ \bibnamefont {Fuller}}, \bibinfo {author}
  {\bibfnamefont {J.~E.}\ \bibnamefont {Davoren}}, \bibinfo {author}
  {\bibfnamefont {P.}~\bibnamefont {Wipf}}, \bibinfo {author} {\bibfnamefont
  {J.~W.}\ \bibnamefont {Kelly}}, \ and\ \bibinfo {author} {\bibfnamefont
  {M.}~\bibnamefont {Gruebele}},\ }\href@noop {} {\bibfield  {journal}
  {\bibinfo  {journal} {Proc Natl Acad Sci USA}\ }\textbf {\bibinfo {volume}
  {105}},\ \bibinfo {pages} {2369} (\bibinfo {year} {2008})}\BibitemShut
  {NoStop}%
\bibitem [{\citenamefont {Albert}\ and\ \citenamefont
  {Barabasi}(2002)}]{Albert:2002wu}%
  \BibitemOpen
  \bibfield  {author} {\bibinfo {author} {\bibfnamefont {R.}~\bibnamefont
  {Albert}}\ and\ \bibinfo {author} {\bibfnamefont {A.}~\bibnamefont
  {Barabasi}},\ }\href@noop {} {\bibfield  {journal} {\bibinfo  {journal} {Rev.
  Mod. Phys.}\ }\textbf {\bibinfo {volume} {74}},\ \bibinfo {pages} {47}
  (\bibinfo {year} {2002})}\BibitemShut {NoStop}%
\bibitem [{\citenamefont {Newman}(2003)}]{Newman:610163}%
  \BibitemOpen
  \bibfield  {author} {\bibinfo {author} {\bibfnamefont {M.~E.~J.}\
  \bibnamefont {Newman}},\ }\href@noop {} {\bibfield  {journal} {\bibinfo
  {journal} {Comput. Phys. Commun.}\ }\textbf {\bibinfo {volume} {147}},\
  \bibinfo {pages} {40} (\bibinfo {year} {2003})}\BibitemShut {NoStop}%
\bibitem [{Note1()}]{Note1}%
  \BibitemOpen
  \bibinfo {note} {Complexity theory has not traditionally focused on this
  topic, but the tools are there for its investigation. See, {\protect \it
  e.g.} \cite {Samukhin:2008hb, Chung:2003el}.}\BibitemShut {Stop}%
\bibitem [{Note2()}]{Note2}%
  \BibitemOpen
  \bibinfo {note} {It should be noted that we do not wish to show this converse
  holds in general. We suspect that there exist systems that are
  thermodynamically two-state, but under certain conditions will display
  multi-exponential kinetics.}\BibitemShut {Stop}%
\bibitem [{\citenamefont {Schindler}\ \emph {et~al.}(1995)\citenamefont
  {Schindler}, \citenamefont {Herrler}, \citenamefont {Marahiel},\ and\
  \citenamefont {Schmid}}]{Schindler:1995wr}%
  \BibitemOpen
  \bibfield  {author} {\bibinfo {author} {\bibfnamefont {T.}~\bibnamefont
  {Schindler}}, \bibinfo {author} {\bibfnamefont {M.}~\bibnamefont {Herrler}},
  \bibinfo {author} {\bibfnamefont {M.~A.}\ \bibnamefont {Marahiel}}, \ and\
  \bibinfo {author} {\bibfnamefont {F.~X.}\ \bibnamefont {Schmid}},\
  }\href@noop {} {\bibfield  {journal} {\bibinfo  {journal} {Nat. Struct.
  Biol.}\ }\textbf {\bibinfo {volume} {2}},\ \bibinfo {pages} {663} (\bibinfo
  {year} {1995})}\BibitemShut {NoStop}%
\bibitem [{\citenamefont {Jacob}\ \emph {et~al.}(1997)\citenamefont {Jacob},
  \citenamefont {Schindler}, \citenamefont {Balbach},\ and\ \citenamefont
  {Schmid}}]{Jacob:1997tw}%
  \BibitemOpen
  \bibfield  {author} {\bibinfo {author} {\bibfnamefont {M.}~\bibnamefont
  {Jacob}}, \bibinfo {author} {\bibfnamefont {T.}~\bibnamefont {Schindler}},
  \bibinfo {author} {\bibfnamefont {J.}~\bibnamefont {Balbach}}, \ and\
  \bibinfo {author} {\bibfnamefont {F.~X.}\ \bibnamefont {Schmid}},\
  }\href@noop {} {\bibfield  {journal} {\bibinfo  {journal} {Proceedings of the
  National Academy of Sciences}\ }\textbf {\bibinfo {volume} {94}},\ \bibinfo
  {pages} {5622} (\bibinfo {year} {1997})}\BibitemShut {NoStop}%
\bibitem [{\citenamefont {Sali}\ \emph {et~al.}(1994)\citenamefont {Sali},
  \citenamefont {Shakhnovich},\ and\ \citenamefont {Karplus}}]{Sali:1994vs}%
  \BibitemOpen
  \bibfield  {author} {\bibinfo {author} {\bibfnamefont {A.}~\bibnamefont
  {Sali}}, \bibinfo {author} {\bibfnamefont {E.}~\bibnamefont {Shakhnovich}}, \
  and\ \bibinfo {author} {\bibfnamefont {M.}~\bibnamefont {Karplus}},\
  }\href@noop {} {\bibfield  {journal} {\bibinfo  {journal} {Nature}\ }\textbf
  {\bibinfo {volume} {369}},\ \bibinfo {pages} {248} (\bibinfo {year}
  {1994})}\BibitemShut {NoStop}%
\bibitem [{Note3()}]{Note3}%
  \BibitemOpen
  \bibinfo {note} {One additional possibility has been suggested by Robert
  McGibbon: it could be that two-state systems are considered ``typical'' by
  the folding community, and non-two state systems are thought of as
  pathological and therefore not studied. For the purposes of this manuscript
  we ignore this possibility, but it is important that those working in the
  field keep it in mind.}\BibitemShut {Stop}%
\bibitem [{\citenamefont {Bowman}\ \emph {et~al.}(2011)\citenamefont {Bowman},
  \citenamefont {Voelz},\ and\ \citenamefont {Pande}}]{Bowman:2010hm}%
  \BibitemOpen
  \bibfield  {author} {\bibinfo {author} {\bibfnamefont {G.~R.}\ \bibnamefont
  {Bowman}}, \bibinfo {author} {\bibfnamefont {V.~A.}\ \bibnamefont {Voelz}}, \
  and\ \bibinfo {author} {\bibfnamefont {V.~S.}\ \bibnamefont {Pande}},\
  }\href@noop {} {\bibfield  {journal} {\bibinfo  {journal} {J. Am. Chem.
  Soc.}\ }\textbf {\bibinfo {volume} {133}},\ \bibinfo {pages} {664} (\bibinfo
  {year} {2011})}\BibitemShut {NoStop}%
\bibitem [{\citenamefont {Pande}(2010{\natexlab{a}})}]{Pande:2010hwa}%
  \BibitemOpen
  \bibfield  {author} {\bibinfo {author} {\bibfnamefont {V.~S.}\ \bibnamefont
  {Pande}},\ }\href@noop {} {\bibfield  {journal} {\bibinfo  {journal} {Phys.
  Rev. Lett.}\ }\textbf {\bibinfo {volume} {105}},\ \bibinfo {pages} {198101}
  (\bibinfo {year} {2010}{\natexlab{a}})}\BibitemShut {NoStop}%
\bibitem [{\citenamefont {Lane}\ and\ \citenamefont
  {Pande}(2012)}]{Lane:2012ba}%
  \BibitemOpen
  \bibfield  {author} {\bibinfo {author} {\bibfnamefont {T.~J.}\ \bibnamefont
  {Lane}}\ and\ \bibinfo {author} {\bibfnamefont {V.~S.}\ \bibnamefont
  {Pande}},\ }\href@noop {} {\bibfield  {journal} {\bibinfo  {journal} {J.
  Phys. Chem. B}\ }\textbf {\bibinfo {volume} {116}},\ \bibinfo {pages} {6764}
  (\bibinfo {year} {2012})}\BibitemShut {NoStop}%
\bibitem [{\citenamefont {Dickson}\ and\ \citenamefont
  {Brooks}(2012)}]{Dickson:2012et}%
  \BibitemOpen
  \bibfield  {author} {\bibinfo {author} {\bibfnamefont {A.}~\bibnamefont
  {Dickson}}\ and\ \bibinfo {author} {\bibfnamefont {C.~L.}\ \bibnamefont
  {Brooks}},\ }\href@noop {} {\bibfield  {journal} {\bibinfo  {journal} {J.
  Chem. Theory Comput.}\ } (\bibinfo {year} {2012})}\BibitemShut {NoStop}%
\bibitem [{\citenamefont {Bowman}\ and\ \citenamefont
  {Pande}(2010)}]{Bowman:2010fp}%
  \BibitemOpen
  \bibfield  {author} {\bibinfo {author} {\bibfnamefont {G.~R.}\ \bibnamefont
  {Bowman}}\ and\ \bibinfo {author} {\bibfnamefont {V.~S.}\ \bibnamefont
  {Pande}},\ }\href@noop {} {\bibfield  {journal} {\bibinfo  {journal} {Proc
  Natl Acad Sci USA}\ }\textbf {\bibinfo {volume} {107}},\ \bibinfo {pages}
  {10890} (\bibinfo {year} {2010})}\BibitemShut {NoStop}%
\bibitem [{\citenamefont {Bogatyreva}\ \emph {et~al.}(2009)\citenamefont
  {Bogatyreva}, \citenamefont {Osypov},\ and\ \citenamefont
  {Ivankov}}]{Bogatyreva:2009jz}%
  \BibitemOpen
  \bibfield  {author} {\bibinfo {author} {\bibfnamefont {N.~S.}\ \bibnamefont
  {Bogatyreva}}, \bibinfo {author} {\bibfnamefont {A.~A.}\ \bibnamefont
  {Osypov}}, \ and\ \bibinfo {author} {\bibfnamefont {D.~N.}\ \bibnamefont
  {Ivankov}},\ }\href@noop {} {\bibfield  {journal} {\bibinfo  {journal}
  {Nucleic Acids Research}\ }\textbf {\bibinfo {volume} {37}},\ \bibinfo
  {pages} {D342} (\bibinfo {year} {2009})}\BibitemShut {NoStop}%
\bibitem [{\citenamefont {Jaynes}(1980)}]{JAYNES:1980uh}%
  \BibitemOpen
  \bibfield  {author} {\bibinfo {author} {\bibfnamefont {E.~T.}\ \bibnamefont
  {Jaynes}},\ }\href@noop {} {\bibfield  {journal} {\bibinfo  {journal} {Annu.
  Rev. Phys. Chem.}\ }\textbf {\bibinfo {volume} {31}},\ \bibinfo {pages} {579}
  (\bibinfo {year} {1980})}\BibitemShut {NoStop}%
\bibitem [{\citenamefont {Stock}\ \emph {et~al.}(2008)\citenamefont {Stock},
  \citenamefont {Ghosh},\ and\ \citenamefont {Dill}}]{Stock:2008ii}%
  \BibitemOpen
  \bibfield  {author} {\bibinfo {author} {\bibfnamefont {G.}~\bibnamefont
  {Stock}}, \bibinfo {author} {\bibfnamefont {K.}~\bibnamefont {Ghosh}}, \ and\
  \bibinfo {author} {\bibfnamefont {K.~A.}\ \bibnamefont {Dill}},\ }\href@noop
  {} {\bibfield  {journal} {\bibinfo  {journal} {J. Chem. Phys.}\ }\textbf
  {\bibinfo {volume} {128}},\ \bibinfo {pages} {194102} (\bibinfo {year}
  {2008})}\BibitemShut {NoStop}%
\bibitem [{\citenamefont {Li}\ \emph {et~al.}(2009)\citenamefont {Li},
  \citenamefont {Oliva}, \citenamefont {Naganathan},\ and\ \citenamefont
  {Mu{\~n}oz}}]{Li:2009bz}%
  \BibitemOpen
  \bibfield  {author} {\bibinfo {author} {\bibfnamefont {P.}~\bibnamefont
  {Li}}, \bibinfo {author} {\bibfnamefont {F.~Y.}\ \bibnamefont {Oliva}},
  \bibinfo {author} {\bibfnamefont {A.~N.}\ \bibnamefont {Naganathan}}, \ and\
  \bibinfo {author} {\bibfnamefont {V.}~\bibnamefont {Mu{\~n}oz}},\ }\href@noop
  {} {\bibfield  {journal} {\bibinfo  {journal} {P Natl Acad Sci Usa}\ }\textbf
  {\bibinfo {volume} {106}},\ \bibinfo {pages} {103} (\bibinfo {year}
  {2009})}\BibitemShut {NoStop}%
\bibitem [{\citenamefont {Lane}\ and\ \citenamefont
  {Pande}(2013)}]{Lane:2013ws}%
  \BibitemOpen
  \bibfield  {author} {\bibinfo {author} {\bibfnamefont {T.~J.}\ \bibnamefont
  {Lane}}\ and\ \bibinfo {author} {\bibfnamefont {V.~S.}\ \bibnamefont
  {Pande}},\ }\href@noop {} {\bibfield  {journal} {\bibinfo  {journal} {arXiv}\
  } (\bibinfo {year} {2013})},\ \Eprint {http://arxiv.org/abs/1301.4302v1}
  {1301.4302v1} \BibitemShut {NoStop}%
\bibitem [{\citenamefont {Prigozhin}\ and\ \citenamefont
  {Gruebele}(2013)}]{Prigozhin:2013ex}%
  \BibitemOpen
  \bibfield  {author} {\bibinfo {author} {\bibfnamefont {M.~B.}\ \bibnamefont
  {Prigozhin}}\ and\ \bibinfo {author} {\bibfnamefont {M.}~\bibnamefont
  {Gruebele}},\ }\href@noop {} {\bibfield  {journal} {\bibinfo  {journal}
  {Phys. Chem. Chem. Phys.}\ } (\bibinfo {year} {2013})}\BibitemShut {NoStop}%
\bibitem [{\citenamefont {Wang}\ \emph {et~al.}(2003)\citenamefont {Wang},
  \citenamefont {Tang}, \citenamefont {Sato}, \citenamefont {Vugmeyster},
  \citenamefont {McKnight},\ and\ \citenamefont {Raleigh}}]{Wang:2003ha}%
  \BibitemOpen
  \bibfield  {author} {\bibinfo {author} {\bibfnamefont {M.}~\bibnamefont
  {Wang}}, \bibinfo {author} {\bibfnamefont {Y.}~\bibnamefont {Tang}}, \bibinfo
  {author} {\bibfnamefont {S.}~\bibnamefont {Sato}}, \bibinfo {author}
  {\bibfnamefont {L.}~\bibnamefont {Vugmeyster}}, \bibinfo {author}
  {\bibfnamefont {C.~J.}\ \bibnamefont {McKnight}}, \ and\ \bibinfo {author}
  {\bibfnamefont {D.~P.}\ \bibnamefont {Raleigh}},\ }\href@noop {} {\bibfield
  {journal} {\bibinfo  {journal} {J. Am. Chem. Soc.}\ }\textbf {\bibinfo
  {volume} {125}},\ \bibinfo {pages} {6032} (\bibinfo {year}
  {2003})}\BibitemShut {NoStop}%
\bibitem [{\citenamefont {Kubelka}\ \emph {et~al.}(2006)\citenamefont
  {Kubelka}, \citenamefont {Chiu}, \citenamefont {Davies}, \citenamefont
  {Eaton},\ and\ \citenamefont {Hofrichter}}]{Kubelka:2006bv}%
  \BibitemOpen
  \bibfield  {author} {\bibinfo {author} {\bibfnamefont {J.}~\bibnamefont
  {Kubelka}}, \bibinfo {author} {\bibfnamefont {T.~K.}\ \bibnamefont {Chiu}},
  \bibinfo {author} {\bibfnamefont {D.~R.}\ \bibnamefont {Davies}}, \bibinfo
  {author} {\bibfnamefont {W.~A.}\ \bibnamefont {Eaton}}, \ and\ \bibinfo
  {author} {\bibfnamefont {J.}~\bibnamefont {Hofrichter}},\ }\href@noop {}
  {\bibfield  {journal} {\bibinfo  {journal} {J. Mol. Biol.}\ }\textbf
  {\bibinfo {volume} {359}},\ \bibinfo {pages} {546} (\bibinfo {year}
  {2006})}\BibitemShut {NoStop}%
\bibitem [{\citenamefont {Reiner}\ \emph {et~al.}(2010)\citenamefont {Reiner},
  \citenamefont {Henklein},\ and\ \citenamefont {Kiefhaber}}]{Reiner:2010bp}%
  \BibitemOpen
  \bibfield  {author} {\bibinfo {author} {\bibfnamefont {A.}~\bibnamefont
  {Reiner}}, \bibinfo {author} {\bibfnamefont {P.}~\bibnamefont {Henklein}}, \
  and\ \bibinfo {author} {\bibfnamefont {T.}~\bibnamefont {Kiefhaber}},\
  }\href@noop {} {\bibfield  {journal} {\bibinfo  {journal} {Proc Natl Acad Sci
  USA}\ }\textbf {\bibinfo {volume} {107}},\ \bibinfo {pages} {4955} (\bibinfo
  {year} {2010})}\BibitemShut {NoStop}%
\bibitem [{Note4()}]{Note4}%
  \BibitemOpen
  \bibinfo {note} {We have only carried out the perturbation to first order.
  The perturbations we want to consider are small, since large perturbations
  stray too far from the entropy maximum solution and therefore are expected to
  imply work done by biology.}\BibitemShut {Stop}%
\bibitem [{\citenamefont {Mehta}(2004)}]{Mehta:2004wq}%
  \BibitemOpen
  \bibfield  {author} {\bibinfo {author} {\bibfnamefont {M.~L.}\ \bibnamefont
  {Mehta}},\ }\href@noop {} {\emph {\bibinfo {title} {{Random matrices}}}}\
  (\bibinfo  {publisher} {Academic Press},\ \bibinfo {year} {2004})\BibitemShut
  {NoStop}%
\bibitem [{Note5()}]{Note5}%
  \BibitemOpen
  \bibinfo {note} {Because of the spreading of the eigenvalues, this gap will
  be smaller than before. One could calculate the probabilistic difference from
  the Tracy-Widom distribution \cite {Tracy:1996uc}, but we have not undertaken
  this task (yet).}\BibitemShut {Stop}%
\bibitem [{Note6()}]{Note6}%
  \BibitemOpen
  \bibinfo {note} {Note that this expression requires $\epsilon \sim \protect
  \mathcal {O}(1/N)$, which for large state number $N$ makes $\epsilon $ very
  small. The choice of $\epsilon $, however, is arbitrary; it is simply a
  control parameter.}\BibitemShut {Stop}%
\bibitem [{\citenamefont {Rao}\ and\ \citenamefont
  {Caflisch}(2004)}]{Rao:2004ce}%
  \BibitemOpen
  \bibfield  {author} {\bibinfo {author} {\bibfnamefont {F.}~\bibnamefont
  {Rao}}\ and\ \bibinfo {author} {\bibfnamefont {A.}~\bibnamefont {Caflisch}},\
  }\href@noop {} {\bibfield  {journal} {\bibinfo  {journal} {J. Mol. Biol.}\
  }\textbf {\bibinfo {volume} {342}},\ \bibinfo {pages} {299} (\bibinfo {year}
  {2004})}\BibitemShut {NoStop}%
\bibitem [{\citenamefont {Lane}\ \emph {et~al.}(2011)\citenamefont {Lane},
  \citenamefont {Bowman}, \citenamefont {Beauchamp}, \citenamefont {Voelz},\
  and\ \citenamefont {Pande}}]{Lane:2011fp}%
  \BibitemOpen
  \bibfield  {author} {\bibinfo {author} {\bibfnamefont {T.~J.}\ \bibnamefont
  {Lane}}, \bibinfo {author} {\bibfnamefont {G.~R.}\ \bibnamefont {Bowman}},
  \bibinfo {author} {\bibfnamefont {K.}~\bibnamefont {Beauchamp}}, \bibinfo
  {author} {\bibfnamefont {V.~A.}\ \bibnamefont {Voelz}}, \ and\ \bibinfo
  {author} {\bibfnamefont {V.~S.}\ \bibnamefont {Pande}},\ }\href@noop {}
  {\bibfield  {journal} {\bibinfo  {journal} {J. Am. Chem. Soc.}\ }\textbf
  {\bibinfo {volume} {133}},\ \bibinfo {pages} {18413} (\bibinfo {year}
  {2011})}\BibitemShut {NoStop}%
\bibitem [{\citenamefont {Dickson}\ and\ \citenamefont
  {Brooks}(2013)}]{Dickson:2013cf}%
  \BibitemOpen
  \bibfield  {author} {\bibinfo {author} {\bibfnamefont {A.}~\bibnamefont
  {Dickson}}\ and\ \bibinfo {author} {\bibfnamefont {C.~L.}\ \bibnamefont
  {Brooks}, \bibfnamefont {III}},\ }\href@noop {} {\bibfield  {journal}
  {\bibinfo  {journal} {J. Am. Chem. Soc.}\ ,\ \bibinfo {pages}
  {130315062412001}} (\bibinfo {year} {2013})}\BibitemShut {NoStop}%
\bibitem [{\citenamefont {Waldauer}\ \emph {et~al.}(2010)\citenamefont
  {Waldauer}, \citenamefont {Bakajin},\ and\ \citenamefont
  {Lapidus}}]{Waldauer:2010vg}%
  \BibitemOpen
  \bibfield  {author} {\bibinfo {author} {\bibfnamefont {S.}~\bibnamefont
  {Waldauer}}, \bibinfo {author} {\bibfnamefont {O.}~\bibnamefont {Bakajin}}, \
  and\ \bibinfo {author} {\bibfnamefont {L.}~\bibnamefont {Lapidus}},\
  }\href@noop {} {\bibfield  {journal} {\bibinfo  {journal} {Proc Natl Acad Sci
  USA}\ }\textbf {\bibinfo {volume} {107}},\ \bibinfo {pages} {13713} (\bibinfo
  {year} {2010})}\BibitemShut {NoStop}%
\bibitem [{\citenamefont {Pande}(2010{\natexlab{b}})}]{Pande:2010hw}%
  \BibitemOpen
  \bibfield  {author} {\bibinfo {author} {\bibfnamefont {V.}~\bibnamefont
  {Pande}},\ }\href@noop {} {\bibfield  {journal} {\bibinfo  {journal} {Phys.
  Rev. Lett.}\ }\textbf {\bibinfo {volume} {105}} (\bibinfo {year}
  {2010}{\natexlab{b}})}\BibitemShut {NoStop}%
\bibitem [{\citenamefont {Samukhin}\ \emph {et~al.}(2008)\citenamefont
  {Samukhin}, \citenamefont {Dorogovtsev},\ and\ \citenamefont
  {Mendes}}]{Samukhin:2008hb}%
  \BibitemOpen
  \bibfield  {author} {\bibinfo {author} {\bibfnamefont {A.~N.}\ \bibnamefont
  {Samukhin}}, \bibinfo {author} {\bibfnamefont {S.~N.}\ \bibnamefont
  {Dorogovtsev}}, \ and\ \bibinfo {author} {\bibfnamefont {J.~F.~F.}\
  \bibnamefont {Mendes}},\ }\href@noop {} {\bibfield  {journal} {\bibinfo
  {journal} {Phys. Rev. E}\ }\textbf {\bibinfo {volume} {77}} (\bibinfo {year}
  {2008})}\BibitemShut {NoStop}%
\bibitem [{\citenamefont {Chung}\ \emph {et~al.}(2003)\citenamefont {Chung},
  \citenamefont {L{\"u}},\ and\ \citenamefont {Vu}}]{Chung:2003el}%
  \BibitemOpen
  \bibfield  {author} {\bibinfo {author} {\bibfnamefont {F.}~\bibnamefont
  {Chung}}, \bibinfo {author} {\bibfnamefont {L.}~\bibnamefont {L{\"u}}}, \
  and\ \bibinfo {author} {\bibfnamefont {V.}~\bibnamefont {Vu}},\ }\href@noop
  {} {\bibfield  {journal} {\bibinfo  {journal} {Proceedings of the National
  Academy of Sciences}\ }\textbf {\bibinfo {volume} {100}},\ \bibinfo {pages}
  {6313} (\bibinfo {year} {2003})}\BibitemShut {NoStop}%
\bibitem [{\citenamefont {Tracy}\ and\ \citenamefont
  {Widom}(1996)}]{Tracy:1996uc}%
  \BibitemOpen
  \bibfield  {author} {\bibinfo {author} {\bibfnamefont {C.~A.}\ \bibnamefont
  {Tracy}}\ and\ \bibinfo {author} {\bibfnamefont {H.}~\bibnamefont {Widom}},\
  }\href@noop {} {\bibfield  {journal} {\bibinfo  {journal} {Communications in
  Mathematical Physics}\ }\textbf {\bibinfo {volume} {177}},\ \bibinfo {pages}
  {727} (\bibinfo {year} {1996})}\BibitemShut {NoStop}%
\end{thebibliography}%

\end{document}